\documentclass[a4paper]{aa}
\usepackage{graphicx}
\usepackage{natbib}

\begin{document}

\newcommand{\Martin}{Mart\'{i}n~}
\newcommand{\Magazzu}{Magazz\'{u}~}

\newcommand{\ltt}{$\stackrel{<}{_\sim}$~}
\newcommand{\CO}{$^{12}$C$^{16}$O~}
\newcommand{\CCO}{$^{13}$C$^{16}$O~}
\newcommand{\vsini}{$v*{\rm sin} i~$}
\newcommand{\mum}{$\mu$m~}

\newcommand{\Vm}{$V_{\rm macro}$}
\newcommand{\Vt}{$V_{\rm t}$}
\newcommand{\Vexp}{$V_{\rm exp}$}
\newcommand{\Vmacro}{$V_{\rm macro}$}
\newcommand{\Vr}{$V_{\rm rotat}$}

\newcommand{\HHO}{H$_2$O~}
\newcommand{\DG}{$\Delta_t$}

\newcommand{\Tef}{\mbox{$T_{\rm eff}~$}}
\newcommand{\Msun}{\mbox{\,$M_{\odot}$~}}
\newcommand{\Lsun}{\mbox{\,$L_{\odot}$~}}
\newcommand{\vunit}{\mbox{\,km\,s$^{-1}$}}

\title{Spectral energy distribution for GJ406}

\author{Ya. V. Pavlenko\inst{1,2}, H.R.A.Jones\inst{1}, Yu. Lyubchik\inst{2},
 J.Tennyson\inst{3}, D.J. Pinfield\inst{1}
}
\institute{Centre for Astrophysics Research, University of Hertfordshire,
College Lane, Hatfield, Hertfordshiere AL10 9AB, UK
\and Main Astronomical Observatory, Academy of Sciences of the Ukraine, Golosiiv
     Woods, Kyiv-127, 03680 Ukraine
\and Department of Physics and Astronomy, University College London,
      Gower Street, London WC1E 6BT UK
}

\offprints{Ya. V. Pavlenko}
\mail{yp@star.herts.ac.uk}

\date{Version of 17 Dec 2003}

\authorrunning{Pavlenko et al.}
\titlerunning{SEDs of GJ406}

\abstract{We present results of modelling the bulk of the spectral
energy distribution (0.35 - 5 \mum) for GJ406 (M6V). Synthetic
spectra were calculated using the NextGen Dusty and Cond model
atmospheres and incorporate line lists for H$_2$O, TiO, CrH, FeH,
CO, MgH molecules as well as the VALD line list of atomic lines.
A
comparison of synthetic and observed spectra gives \Tef = 2800
$\pm$ 100 K. We determine  M$_{\rm bol}$ = 12.13 $\pm$ 0.10
for which evolutionary
models by Baraffe et al. (2003) suggest an age of around  0.1
-- 0.35 Gyr consistent with its high activity. The age and
luminosity of GJ406 correspond to a wide range of plausible masses
(0.07 -- 0.1 \Msun). 
}
\maketitle

\section{Introduction}

Studies of M dwarf spectra are of interest to many branches of
modern astrophysics. Indeed, perhaps 70 \% of stars within 10
parsecs are M dwarfs and it is very probable that this number
density prevails throughout our Galaxy. The population of these
numerous low-mass stars (0.08 \Msun \ltt $M$ \ltt 0.6 \Msun),
together with substellar objects (brown dwarfs; M $ \leq $ 0.075
\Msun) would contain an appreciable amount of the baryonic matter
in the Galaxy. Estimates of brown dwarf number densities currently
suggest the same order as for stars ($\sim$ 0.1 per pc$^{-3}$),
therefore their contribution to the total mass should not exceed
15 \% (Reid et al. 1999). Nonetheless, the large errors associated
with age and mass determinations for brown dwarfs make such
estimates very uncertain.

The verification of the theory of stellar evolution and structure of
stars, the detection among M dwarfs of a subset of young brown dwarfs,
and the physical state of plasma in their low temperature atmospheres
are among a few of the interesting problems that may be addressed
through the detailed study of M-dwarfs.

Some  authors reference
GJ406  (other names are V*CN Leo, EUVE J1056+07.0, [GKL99] 228,
GSC 00261-00377, LFT 750, LHS 36, 2MASS J10562886+0700527,
 2RE J1056+070, 1RXS J105630.3+070118)
as an ``archetype dwarf of spectral type M6V'', or one of the
``well known spectral standards for its type'' (Mohanty et al.
2004). GJ 406 is located at 2.39 pc from the Sun (Henry et al.
2004). Altena et al. (1995) determined a proper motion of $\mu$ =
4.696. Radial velocities are of order 19 $\pm$ 0.1 km/s (Martin et
al. 1997, Mohanty \& Basri 2003, Fuhrmeister et al. 2005). Leggett
(1992) found that this nearby dwarf has typical old disk
properties. Deflosse et al. (1998) reported a rather low \vsini
$<$ 3 km/s (see also Mohanty \& Basri 2003). Guetter et al. (2003)
even use GJ 406 as one of the JHK standard stars on the CIT
photometrical system. Indeed, optical and IR spectra of the dwarf
do not contain any unusual features. They are governed by
absorption of diatomic molecular band systems, such as TiO and VO
(Kirkpatrick, Henry \& McCarthy 1991), as well as
rotational-vibrational bands of  \HHO and CO (Jones et al. 1994).

On the other hand, GJ 406 has some very unusual properties.
SIMBAD (http://simbad.u-strasbg.fr/sim-fid.pl) labels GJ 406 as a flare star.
The dwarf has strong H$_{\alpha}$ of EW = 6.7 \AA ($L_{H_{\alpha}}/
L_{\rm bol} = -3.9$, see Mohanty \& Basri 2003).
An ultraviolet spectrum of GJ406 contains some emission lines (Fuhrmeister et al.
2004). Furthermore, Schmitt \& Wichmann (2001) detected the Fe XIII forbidden
coronal line at 0.33881 \mum. Fuhrmeister at al. (2004) reported
a high level of variability of this line on a timescale of hours which they
ascribe to microflare heating.  Recently an X-ray luminosity of log $L_x$ = 26.97
was detected by Schmitt \& Liefke (2004). GJ 406 is the only known
M6 star yet observed with a strong chromospheric and
coronal activity. Only a few stars in the solar vicinity are known with such a
menagerie of activity phenomena.

In this paper we compute the synthetic energy distribution of
several model atmospheres with a range of effective temperature
and compared them with the observed fluxes of GJ406. Section 2
presents the spectral data used in our paper. Section 3 describes
our procedure for computation. We psesent our results in Section
4. In Section 5 we discuss the implications of our results.

%
%

\section{Observations}

\begin{table*}
\caption{\label{__obs} Telescope and Instrument configurations used to collect
our GJ 406 dataset.}
  \begin{tabular}{lllll}
Start & End & Instrument (configuration) & Telescope &Date\\
\mum  & \mum &                           &           &    \\
\hline
\hline
\noalign{\smallskip}
0.35 & 0.56  &ISIS (blue arm) & WHT & 2001 Jan 29 \\
0.55 & 0.80  &ISIS (red arm) & WHT & 2001 Jan 29 \\
0.79 & 1.20  &NICMOS (G096) & HST & 1998 June 19\\
1.05 & 1.95  &NICMOS (G141) & HST & 1998 June 19\\
1.3 & 2.59  &NICMOS (G206) & HST & 1998 June 19\\
2.48 & 2.60  &SWS (06 1A) & ISO & 1996 June 26 \\
2.60 & 2.75  &SWS (06 1A) & ISO & 1996 June 26 \\
2.74 & 2.90  &SWS (06 1A) & ISO & 1996 June 26 \\
2.88 & 3.02  &SWS (06 1B) & ISO & 1996 June 26 \\
3.03 & 3.23  &CGS4 (150 l/mm) & UKIRT & 1993 April 20 \\
3.21 & 3.40  &CGS4 (150 l/mm)& UKIRT & 1993 April 20 \\
3.36 & 3.75  &CGS4 (75 l/mm) & UKIRT & 1992 May 7 \\
3.76 & 4.15  &CGS4 (75 l/mm) & UKIRT & 1992 May 7 \\
4.51 & 4.90  &CGS4 (75 l/mm) & UKIRT & 1992 October 26 \\
\noalign{\smallskip}
\hline

\end{tabular}
\end{table*}

\begin{figure*}
\begin{center}
\includegraphics [width=140mm]{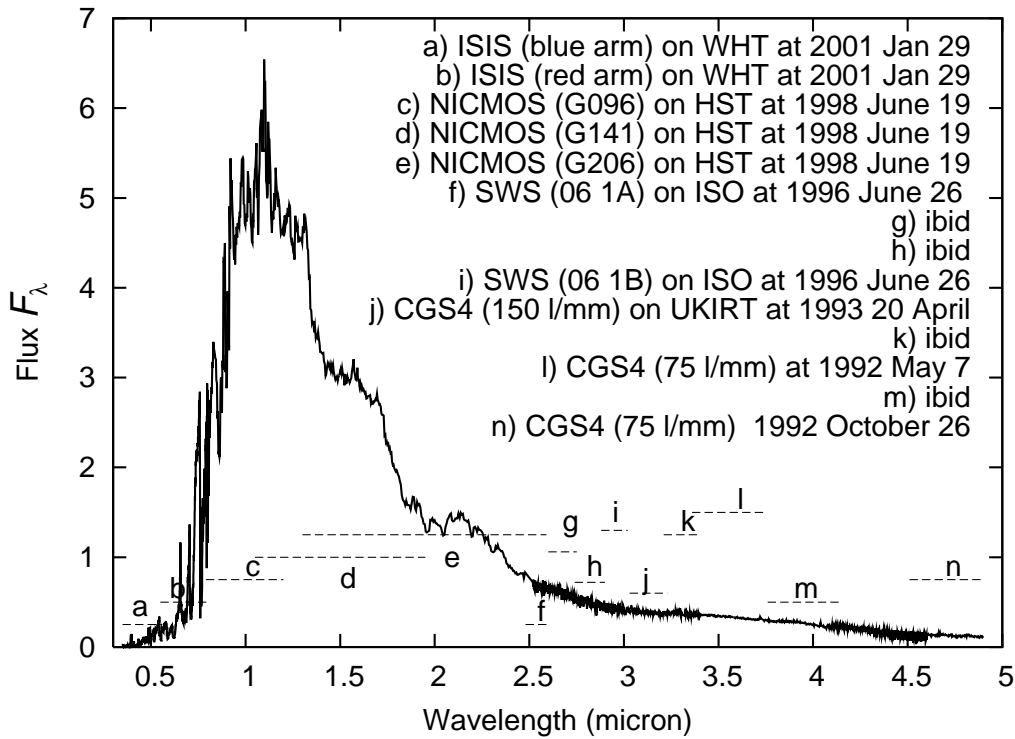}
\end{center}
\caption[]{\label{__obs_data} Observational data used for this
paper. The wavelength coverage of the different instruments is
shown.}
\end{figure*}

Table \ref{__obs} lists the data and instruments
used to obtain the observational spectra in this paper.
The spectra are shown in Fig. \ref{__obs_data} and
come from measurements taken
with a variety of different instruments on different telescopes.
All the data are assessed to be of good quality, most have already
be used for other papers. We refer to the spectra in wavelength order.
The reduction procedures for the Integral Spectrometer (ISIS) data taken on the William Herschel
Telescope (WHT) are described in Dobbie et al. (2004). The
reduction procedures for
the CGS4 data taken on the United Kingdom Infrared Telescope (UKIRT)
here have been reported elsewhere in Jones et
al. (1994, 1995, 1996). It should be noted that CGS4 observations in the spectral regions labelled
in Fig. \ref{__obs_data} by ``m'' and ``n'' are not continuous.
Fluxes between them were filled by
NextGen synthetic spectra.
 The reduction procedures for the Short Wavelength
spectrometer (SWS) data taken with the Infrared Space Observatory
(ISO) are reported elsewhere in Jones et al. (2002).  The Near
Infrared Camera Mosiac Spectrograph (NICMOS) data taken on the
Hubble Space Telescope (HST) was reduced using the data processing
software Calnicc (Version 2.5.7). The data has been compared with
Jones et al. (1994) and is preferred due to its excellent flux
calibration (5-10\%, Pirzkal \& Freudling 1998) and broad
wavelength coverage.

In general we have renormalised the fluxes for the different
spectral regions so as to ensure that flux levels are the same
where the regions overlap. For the CGS4 spectra centered on 4.7
\mum (region n), this was not possible since there is no overlap.
For this spectral region, we simply used the instrument
calibration to determine the flux level, and filled in the gap
with a normalised section of NextGen model spectra.

\subsection{Absolute flux calibration.}
In order to provide an absolute normalisation for the full GJ406
spectrum, we used the available near infrared photometry from
Leggett et al. (2000), which we transformed onto the Mauna Kea
Observatory (MKO) photometric system using Hawarden et al. (2001).
We used the measured MKO J, H and K filter pass bands convolved
with atmospheric transmission to estimate the ground based
photometrically measured flux components in our spectra, by
integrating over each band. We then performed the same task on a
flux calibrated spectrum of Vega. Vega was assumed to be zero
magnitude at all wavelengths, and the GJ406 flux could thus be
scaled to match the photometry. The normalisation value that we
found was $\sim$20\% higher when normalising in the K-band than in
the J-band, and was an intermediate value in the H-band. These
differences presumably result primarily from the relative
normalisations that we used to join the individual spectral
regions, and provide an accuracy gauge on this procedure. We chose
to make our final normalisation in the H-band, and duly estimate a
likely $\pm$10\% uncertainty in our absolute flux levels.

In order to derive a bolometric flux, we added a 4.8--20 \mum
spectral tail to our calibrated spectrum (using a NextGen 2800 K,
$\log{g}$=5.0, [M/H]=0 model spectra, shown to be appropriate in
Section 4.2), normalised in its overlap region, and integrated out
to 20 \mum. Note however, that this synthetic spectral tail only
contributes $\sim$3\% to the bolometric flux, which we found to be
6.35$\times$ 10$^{-12}$ W m$^{-2}$. We then derived the bolometric
magnitude (m$_{\rm bol}$) using the Sun as a standard (adopting
$L_{\odot}$=3.86$\times$ 10$^{26}$ W and M$_{\rm bol\odot}$=4.75),
which yields m$_{\rm bol}$=9.02. This is consistent with the
value derived by Leggett et al. (2000) of 9.07.
Assuming a distance modulus of $m$ = -3.11 $\pm$ 0.01 (van Altena
et al. 1995) and 10\% uncertainty in our flux calibration, we thus
determine that GJ406 has M$_{bol}$=12.13$\pm$0.10 and $L$ =
$\log{L_*/L_{\odot}}$=-2.95$\pm$0.05.


\section{Theoretical spectra computation procedure.}

Theoretical spectral energy distributions\footnote{Hereafter
we use the term ``synthetic spectra'' to simplify
the text.} were computed
for model atmospheres of
dwarfs with effective temperatures T$_{\rm eff}$ = 2500--3200 K from the
NextGen grid of Hauschildt, Allard \& Baron (1999) for solar metallicity
(Anders \& Grevesse 1989). Hereafter we use the syntax ``effective temperature/
gravity/metallicity'', e.g 2800/5.0/0 to signify the model atmosphere.
Unless otherwise
mentioned all models are for log g = 5.0.
Computations of synthetic spectra were carried out by the
program WITA6 (Pavlenko 2000) assuming LTE, hydrostatic
equilibrium for a one-dimensional model atmosphere and without
sources and sinks of energy. The equations of
ionisation-dissociation equilibrium were solved for media
consisting of atoms, ions and molecules. We took into account
$\sim$ 100 components (Pavlenko 2000). The constants for equations
of chemical balance were taken from Tsuji (1973).

Molecular line data were taken from different sources.
Lines of $^1$H$_2^{16}$O were computed using the AMES
database (Partrige \& Schwenke 1998).
The partition functions  of
\HHO ~were also computed from these data (see section 3.1).
 \CO and \CCO ~line lists were computed by Goorvitch (1994).
The CO partition
functions were taken from Gurvitz et al. (1989).
TiO line lists were taken from Plez (1998) and Schwenke (1998).
CN lines came from CDROM 18 (Kurucz 1993);
CrH and FeH lines  were taken from Burrows et al. (2002) and Dulick et
al.(2003), respectively.
Atomic line list was taken from VALD (Kupka et al. 1999).

\begin{figure}
\begin{center}
\includegraphics [width=88mm]{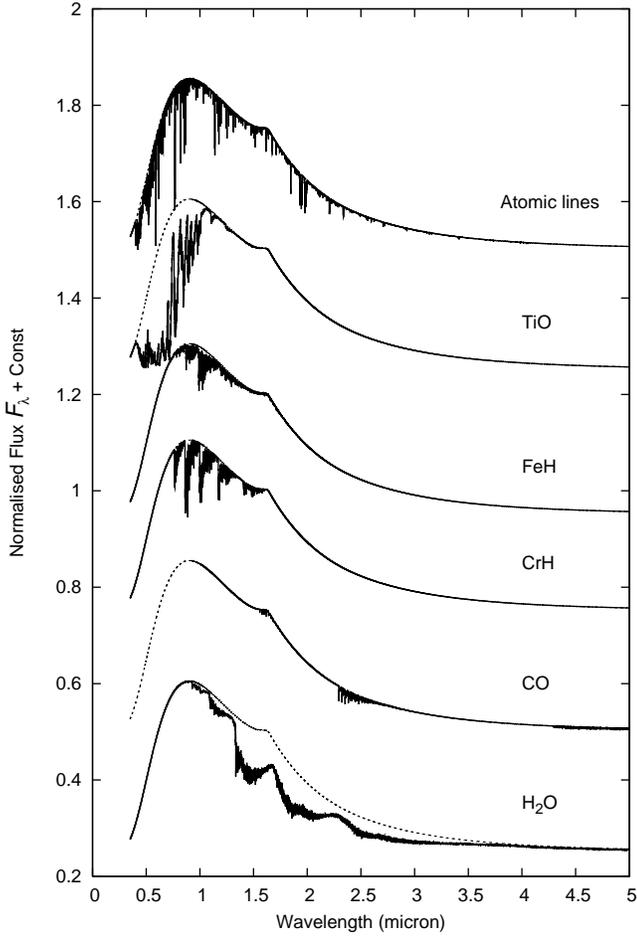}
\end{center}
\caption[]{\label{__opacities} The contribution of different molecules to
the formation of the synthetic spectrum of in \Tef/log g = 2800/5.0 model
atmosphere.}
\end{figure}

The profiles of molecular and atomic lines were determined using
the Voigt function $H(a,v)$. Parameters of their natural
broadening $C_2$ and van der Waals broadening $C_4$ were taken
from Kupka et al. (1999) or in their absence computed following
Uns\"old (1955). Owing to the low temperatures in M dwarf
atmospheres and consequently, electron densities, Stark broadening
could be neglected. As a whole the effects of pressure broadening
prevail. Computations for synthetic spectra were carried out with
a step 0.5 \AA~ for microturbulent velocity $v_t$ = 1 - 4 km/s.
The instrumental broadening was modelled by gaussian profiles set
to approximate the resolution of the observed spectra. The
relative importance of the different opacities contributing to our
synthetic spectra is shown in Fig. \ref{__opacities}.

\subsection{Partition functions of water}

%

We recomputed the constants of chemical equilibrium following
Kurucz (1970) taking into account weights s$_i$ = 1/4 and 3/4 for levels
of water of different symmetry. We followed the scheme described
by Pavlenko (2002). Let us write an equation of
ionisation-dissotiation equilibrium for the molecule consisting of
{\it x, y,...., z}  atoms as

\begin{eqnarray}
n_x*...*n_z/n_{x...z}=  exp(-E_{xy...z}/T_{ev}+b-c* \nonumber \\
(T+d*(T-e*(T+f*T)))+ \nonumber \\ 3/2*(m-k-1)*ln T)
\end{eqnarray}
where $E$ and $T$ are dissociation energy and temperature (in eV), $n_z$
is the number density of z-species, k and m are ionisation degree (0 for
neutrals) and number of atoms per molecule, respectively (see
Kurucz (1970) for more details). Computed constants $a,b,c,d,e,f$
are given in Table \ref{__table_pf}.

\begin{figure}
\begin{center}
\includegraphics [width=88mm]{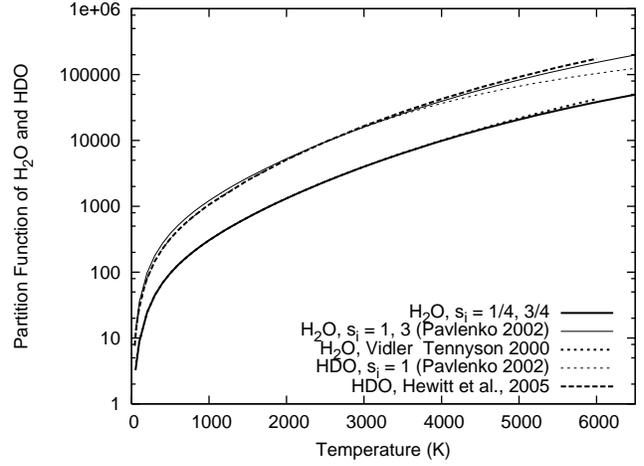}
\end{center}
\caption[]{\label{__pf} Partition functions of molecules \HHO and HDO.
Our data for \HHO are compared with Vidler \& Tennyson(2000), data for
HDO are compared with Hewitt et al. (2005). Our partition
function for HDO is computed for s$_1$ =1. The differences in the
 HDO partition
functions at higher temperatures (T $>$ 4000 K) are due to the use of
more complete sets of deuterated water levels in the UCL model
compared with the AMES model.}
\end{figure}

\begin{figure}
\begin{center}
\includegraphics [width=88mm]{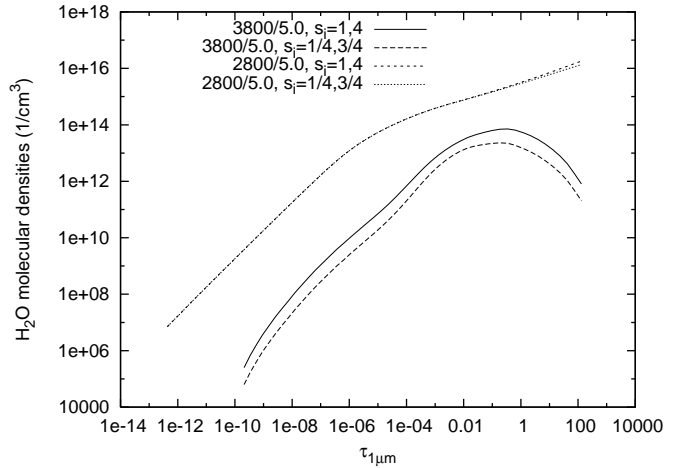}
\end{center}
\caption[]{\label{_ppf} Molecular densities of \HHO computed for
NextGen model atmospheres 2800/5.0 and 3800/5.0 for the cases
of partition functions computed with $s_i$ = 1,4 and $s_i$ =1/4, 3/4.}
\end{figure}

\begin{table*}
\caption{\label{__table_pf} Constants of the dissociation equilibrium of \HHO in formats
of the ATLASxx, where xx labels a version of ASTLAS, the superscripts
correspond to different fit temperature ranges:
$^1$ --- 300 K $<$ T $<$ 1000 K,
$^2$ --- 300 K $<$ T $<$ 6000 K,
$^3$ --- 60 K $<$ T $<$ 6000 K,
$^4$ --- 50 K $<$ T $<$ 10000.
}
\begin{tabular}{|c|c|c|c|c|c|c|}
\hline
\noalign{\smallskip}
 $D_{o}$ & $b$      &  $c$ &        $d$       &  $e$  &     $f$ & Refs. \\
\hline
\noalign{\smallskip}
       &           &           &           &         &      &        \\
 9.500& 0.9193E+02& 0.2550E-02& 0.4088E-06& 0.3893E-10& 0.1512E-14& Pavlenko (2002)$^1$ \\
 9.512& 9.3179E+01& 2.6725E-03& 5.7830E-07& 8.5268E-11& 5.1311E-15& Kurucz (1999) \\
9.500 & 0.9374E+02 & 0.3494E-02&  0.9795E-06&  0.1655E-09&  0.1073E-13 & Vidler \& Tennyson (2000)$^2$ \\
9.500 & 0.9494E+02 & 0.5858E-02&  0.2338E-05&  0.4634E-09&  0.3279E-13 & Vidler \& Tennyson (2000)$^3$ \\
9.500 & 0.9331E+02 & 0.2551E-02&  0.4089E-06&  0.3895E-10&  0.1513E-14 & This work$^1$ \\
9.500 & 0.9428E+02 & 0.3722E-02&  0.8305E-06&  0.9719E-10&  0.4238E-14 & This work$^4$ \\
      &           &           &           &            &          &             \\
\hline
\noalign{\smallskip}
\end{tabular}

\end{table*}

The temperature dependence of computed partition functions of \HHO are given
in Fig. \ref{__pf}. In general, our new partition functions agree well
with data of Vidler \& Tennyson (2000) computed using a mixture of
experimental data and a UCL model of
the water vapour molecule. Some differences occur at T $>$ 5000 K as the UCL
model has more levels of high excitation energy.

It is worth noting that \\
-- the molecular densities of given molecules
obtained from a solution of the system of equations of molecular
equilibrium response to changes of absolute values of $U$,
especially in high temperature regimes (T $>$ 3000 K, see
Fig. \ref{_ppf}).

-- $U$(\HHO) depends strongly on
temperature.
Table \ref{__table_pf} provides fitting constants
obtained for different temperature regions. From a general point of
view it would be reasonable to restrict our fitting to
temperatures T $>$ 300 K. Though for some astrophysical
objects it would be
interesting to have the partition function of water vapour for even
lower temperatures. Thus we have provided these as well.

\section{Results}

\subsection{Dependence of theoretical spectra on input parameters}

\begin{figure*}
\begin{center}
\includegraphics [width=140mm]{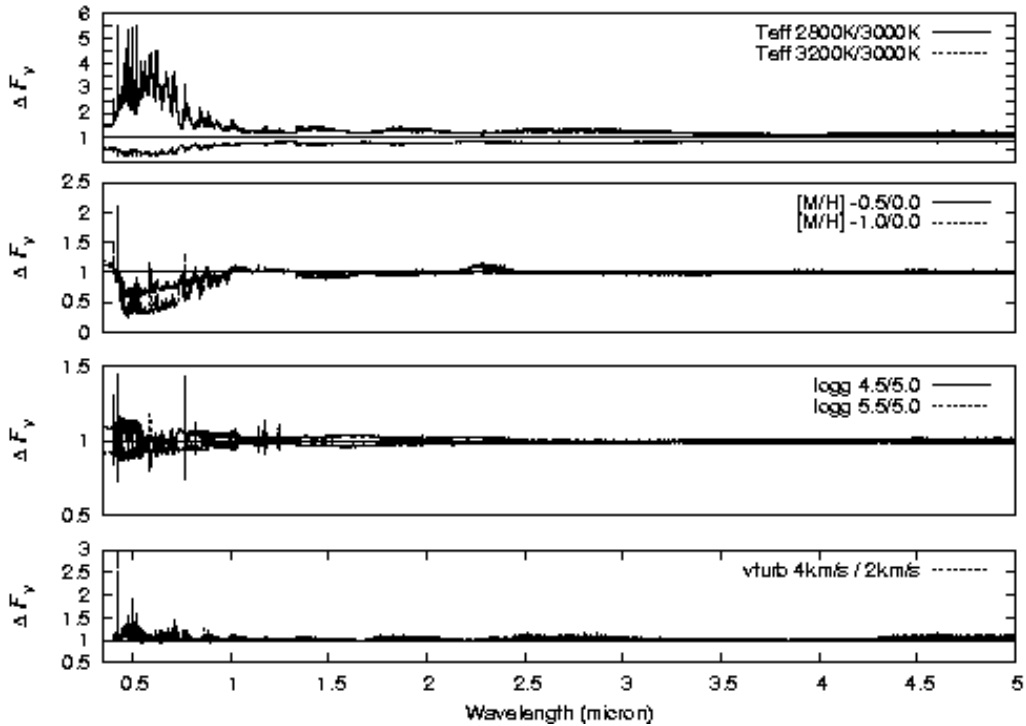}
\end{center}
\caption[]{\label{__deviat} Responses of computed spectra to
variations of input parameters. A model atmosphere of 3000/5.0/0.0
was used as the reference model.}
\end{figure*}

First of all, we computed model SED's using different input
parameters: effective temperatures, gravities, metallicities,
microturbulent velocities. Ratios of fluxes computed with different
sets of input parameters are shown in Fig.
\ref{__deviat}. It can be seen that
a temperature change of 200~K is roughly equivalent to a change in
metallicity of 0.5~dex or a gravity change of $\Delta$log~$g$~=~1.

In Fig. \ref{__deviat}  we see the differential effects of the
dependence of our model spectra on different parameters. Some of
these effects depend more on changes in the opacities for different
parametres. However, some effects are explained by changes in the
structures of the model atmosphere. Indeed, stellar photospheres
of different \Tef, log g, [M/H] lie in different pressure regions.

\begin{figure}
\begin{center}
\includegraphics [width=88mm]{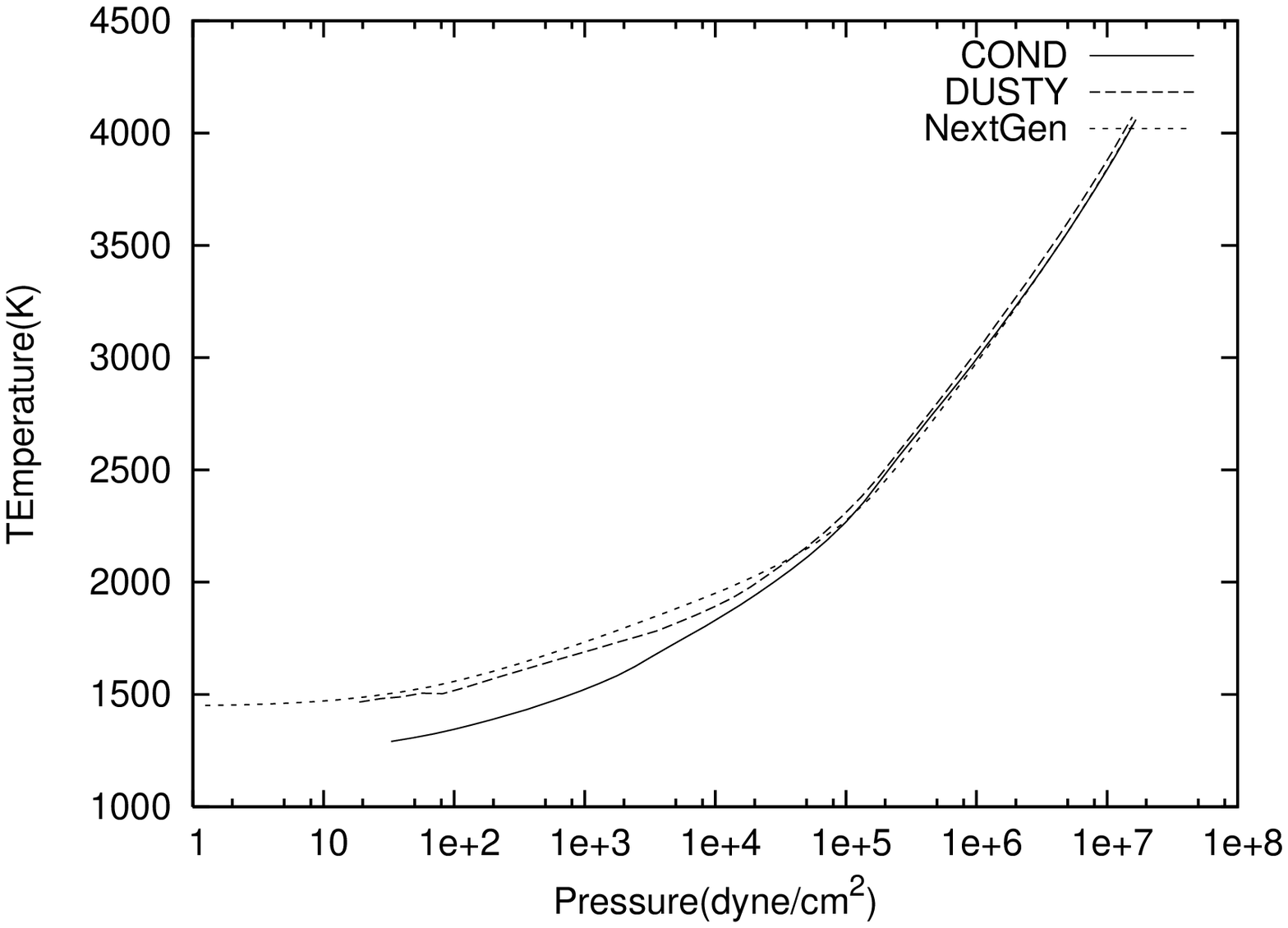}
\includegraphics [width=88mm]{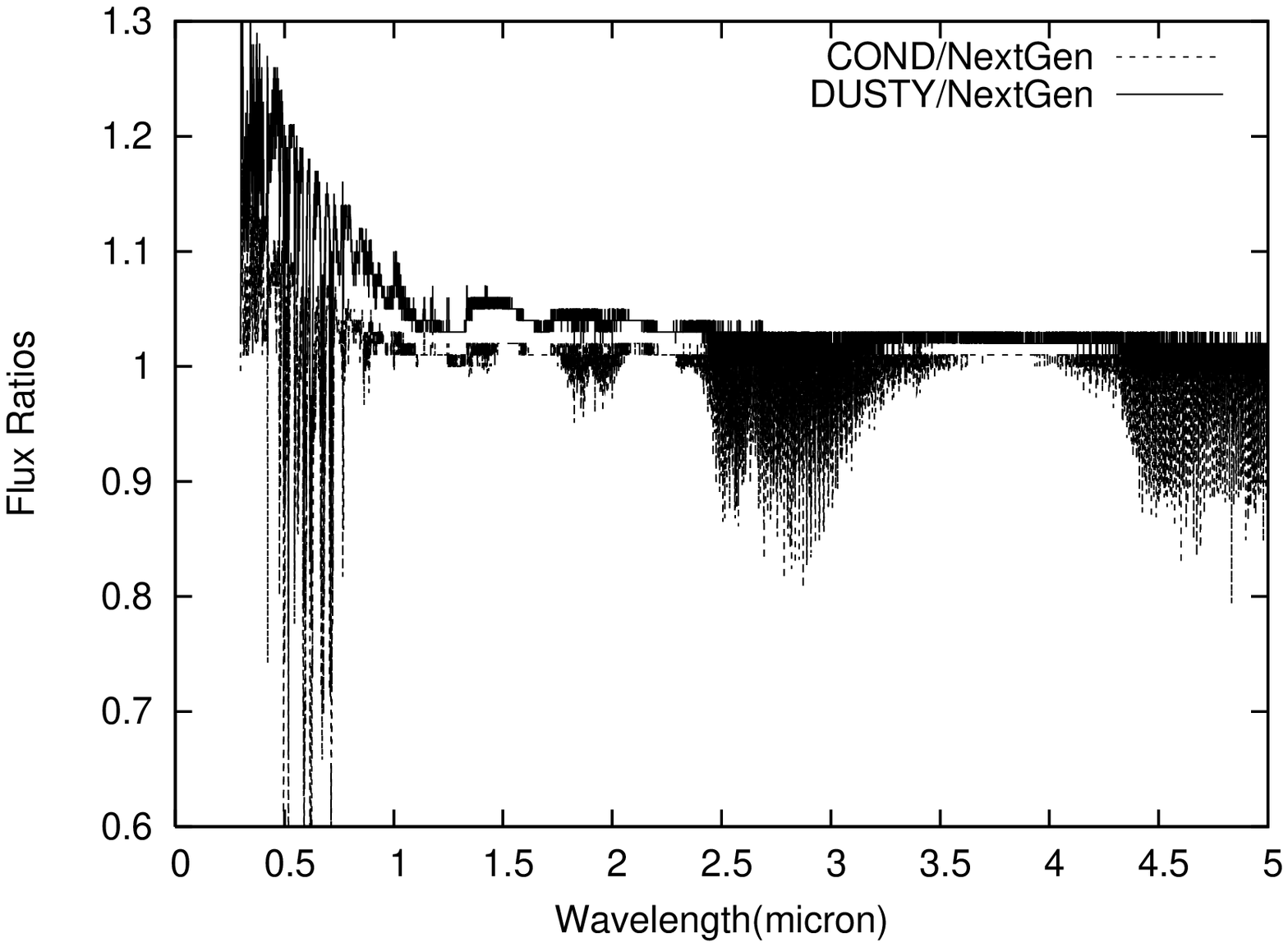}
\end{center}
\caption[]{\label{_df} Top: Temperature structure of model NextGen, DUSTY and COND
model atmospheres of 2800/5.0/0. Bottom: ratio of fluxes computed for these
model atmospheres.}
\end{figure}

\begin{figure}
\begin{center}
\includegraphics [width=88mm]{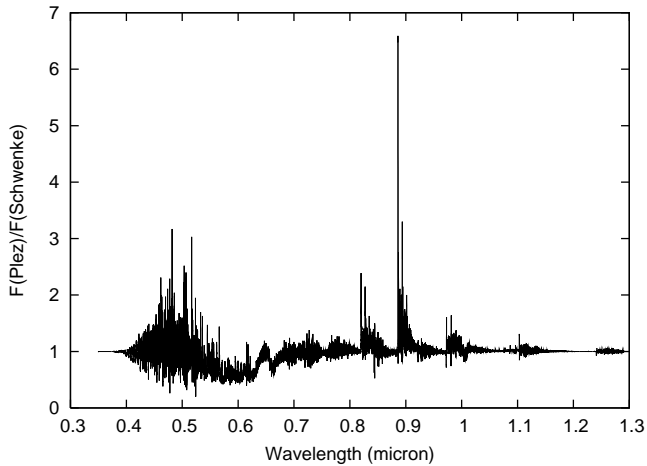}
\end{center}
\caption[]{\label{__TiO} Ratio of fluxes computed for 2800/5.0
NextGen model atmosphere
with TiO line lists by Plez (1998) and Schwenke(1998).}
\end{figure}

\begin{figure}
\begin{center}
\includegraphics [width=88mm]{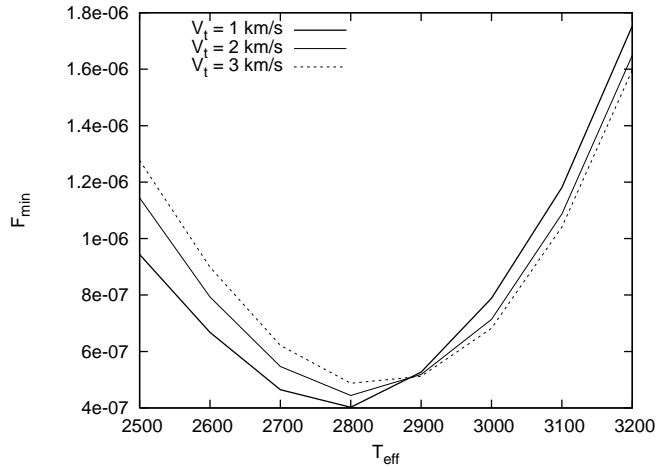}
\end{center}
\caption[]{\label{__x} Min $F(f_s, f_n$) found for different
model atmospheres.}
\end{figure}

\begin{figure*}
\begin{center}
\includegraphics [width=170mm,height=65mm]{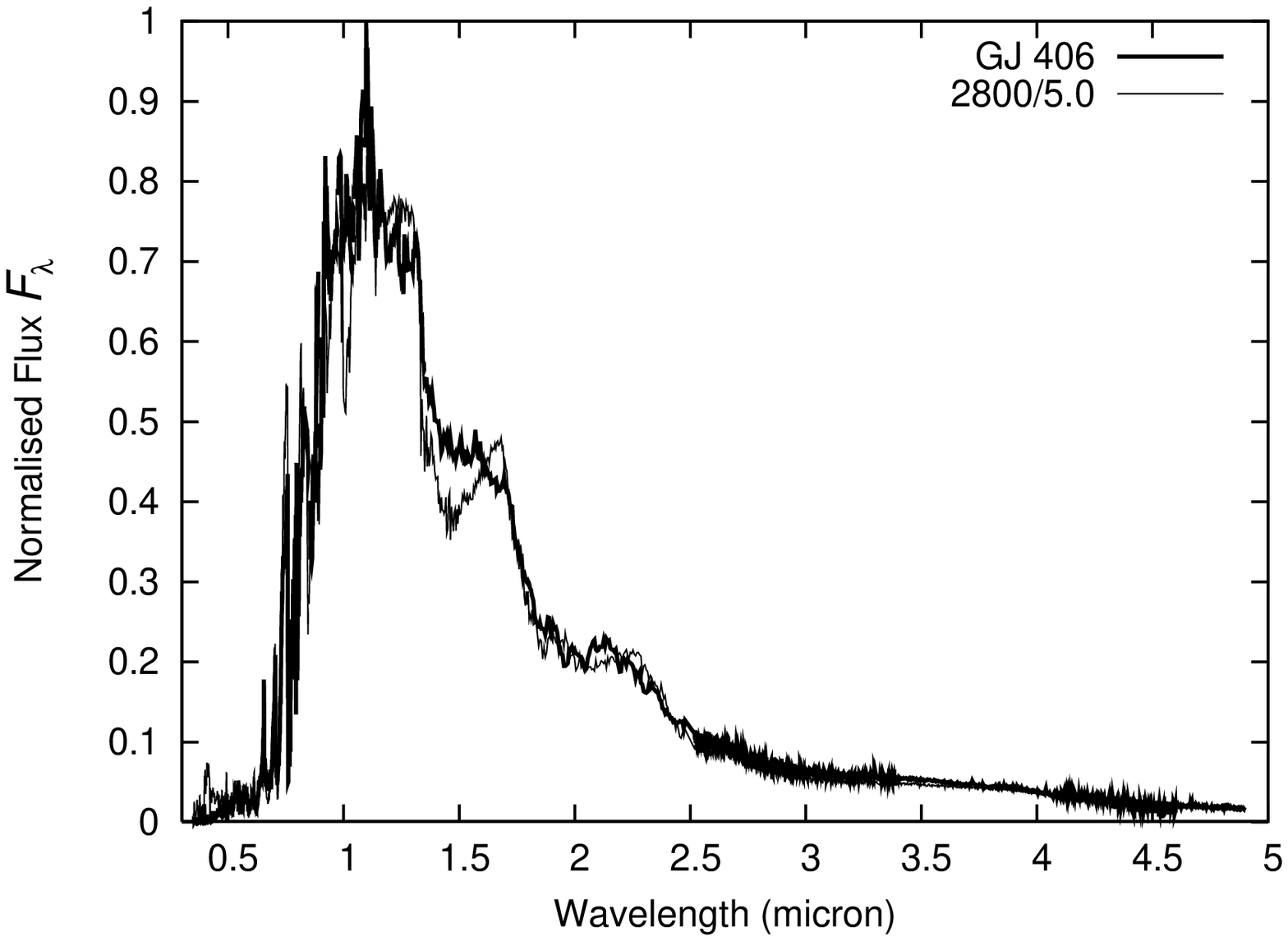}
\includegraphics [width=170mm,height=65mm]{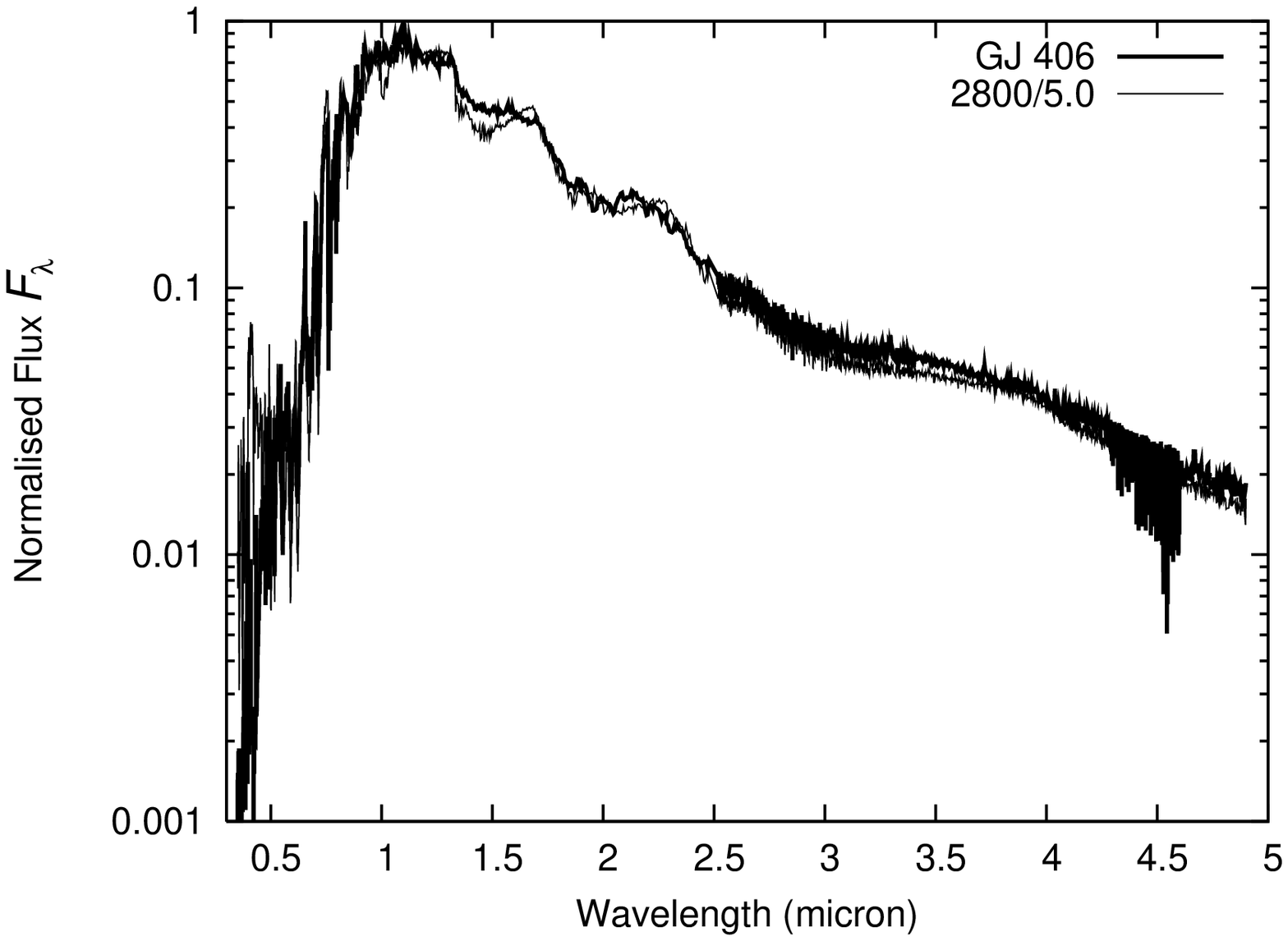}
\end{center}
\caption[]{\label{__fits1} Fits with linear and logarithmic
flux  scales to our GJ406 spectrum with a
theoretical spectrum computed for a solar composition
NextGen model atmosphere 2800/5.0.}
\end{figure*}

\subsection{Dependence of theoretical spectra on different model atmospheres}

We also examined the dependence of our results on the choice of
model atmospheres: NextGen, Dusty, Cond (see Allard 2005 for
references). These model atmospheres have different $T=f(P)$
structures (Fig. \ref{_df}) due to differences in the physical
treatment of the dust formation
 which cause some changes in the opacities and the molecular equilibrium.
In general, the DUSTY and COND models have more hot inner layers
and cooler outer layers (see first plot in Fig. \ref{_df}).

We computed model theoretical fluxes for the NextGen, Dusty and COND
model atmospheres with parameters 2800/5.0/0 in the spectral region of
interest and compared them. Theoretical fluxes were convolved
with a spectral
resolution element of FHWM=4 \AA. In Fig. \ref{_df} we show the ratio of the
convolved fluxes.

It is worth noting a few results:

--- Infrared spectra containing strong water bands agree rather well.

--- We see rather big differences in the optical and blue parts of the spectra.
Here the dependence of computed spectra to changing temperature is
much stronger in comparison to the IR spectral region. In general,
both the COND and DUSTY model atmospheres are a bit hotter in the inner regions,
and  we therefore see an increased flux in the continuum flux levels
(or increased flux for lower opacity wavelength ranges).
Some molecular lines here become stronger due to
the lower temperatures in DUSTY and COND model atmospheres.

--- CO bands strength responds to the change in temperature in the
photospheric layers, due  to the high sensitivity of dissociation
equilibrium of this molecule to temperature.

\subsection{Dependence of theoretical spectra on the use if different TiO line lists.}

There are two TiO line lists of widespread use
by Schwenke(1998) and Plez(1998). They are based on the improved
Langhoff(1997) model of the TiO molecule but differ in details.
Plez (1998) added an a-f system at 0.5 \mum to the line list. Schwenke
subsequently computed a corresponding list of transitions complete to
the higher excitation energies. Plez (1998) provided a line list for a solar mixture of
Ti isotopes. Schwenke's (1998) database provides lists for each
TiO isotope.

We compare synthetic spectra computed with these two line lists for
a NextGen model
atmosphere 2800/5.0/0 (Fig. \ref{__TiO}). Synthetic spectra were
computed with a  0.1 \AA~ step, then convolved with
a spectral resolution element of FWHM = 1 \AA.

 We find the largest differences  in the blue part of spectrum, which is more
affected by incompleteness of molecular line lists for other molecules,
chromospheric effects, veiling, and strong atomic absorption.
Therefore
from inspection of Fig. \ref{__TiO} we conclude that differences
between synthetic spectra computed with Plez (1998) and Schwenke (1998) line
lists do not affect our main results (see also
Lyubchik \& Pavlenko 2001).

\subsection{Fits to GJ406 spectra}

In this paper we are interested in the dependence of
computed spectral energy distributions to \Tef. To find the best
fit of computed spectra to observed fluxes we use a minimisation
procedure described in Jones et al (2002) and Pavlenko \& Jones
(2003). Namely, the best fits are found for the  $min~F(f_s,f_n,
f_g)$, where $f_s$, $f_n$ and $f_g$ are relative shift of the
spectra, normalisation constant for the computed spectra, and
broadening parameter, respectively. We found a rather weak
dependence of the  $min~F$ on the $f_g$ broadening parameter
and set $f_g$ = 6 \AA. Previous studies have considered GJ406 as a
typical M6 dwarf, and we thus assumed log g = 5.0 in its
atmosphere.

To determine a self-consistent solution we  fit the theoretical
spectra to  the observed fluxes in all
spectral regions and estimate the quality of the fit by computing $F(f_s, f_n$).
Two spectral regions were excluded from our
analysis: 0.35 -- 0.4 \mum due
to incompleteness of our opacity sources, and 4.3 -- 0.461 \mum due to the gap
in the observed data.

In Fig. \ref{__x} we show computed $F(f_s, f_n$) for a grid of our
theoretical spectra of different \Tef. We find a weak dependence
of $F(f_s, f_n$) on \Vt. The best fit can be found for the min
$F(f_s, f_n$) at 2800 $\pm$ 100 K ( Fig. \ref{__fits1} shows these
fits in linear and logarithmic scales).

Some problems with fitting spectral features are seen at 1.3
$< \lambda <$ 1.7 \mum. Partially these discrepancies can be
exlained by problems with modelling of strong \HHO bands located
here (Jones at al. 2005).
 Fits to TiO and VO in optical spectral regions and \HHO
(beyond 1.7 \mum) are of particularly good quality (see Pavlenko 1998, Jones
et al. 2002 for more details).

\begin{figure}
\begin{center}
\includegraphics [width=88mm]{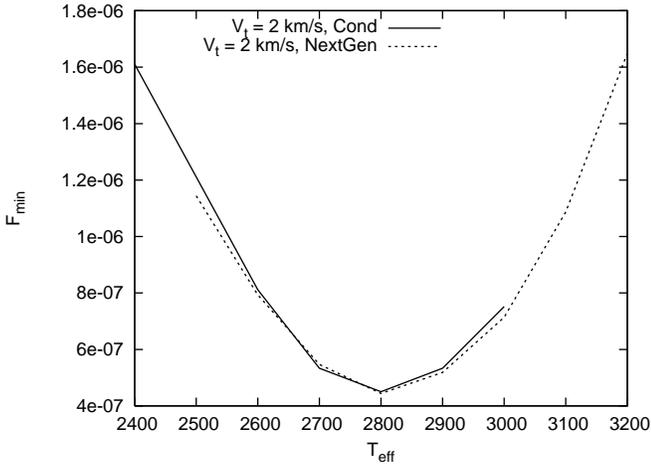}
\end{center}
\caption[]{\label{_fcond} A comparison of $f(x_s, x_n$) computed for the fitting
of onserved spectra by NextGen and Cond synthetic spectra.}
\end{figure}

We also made comparison of the observed spectrum with synthetic
spectra based on the Cond model atmospheres and obtained the fits
practically of the the same quality. The ``best fit'' synthetical
spectra computed with NextGen and Cond model atmospheres coincide
over the whole spectral region. The largest differences do not
exceed 1-2 \%. Consequently, the function $f(x_s, x_n)$ for fits
to the Cond synthetic spectra has the minimum at 2800 K, as for
NextGen (see Fig. \ref{_fcond}).

To confirm this explanation of the discrepancy between theory and observation,
it would be desirable to carry out a similar analysis
of late-type
dwarfs like GJ 406, but employing a more complete water vapor line
list for the
theoretical atmosphere calculations. Such a line list is
expected to be available
in the near future (Jones et al. 2005).

\subsection{Evolutionary model fits}

\begin{figure*}
\begin{center}
\includegraphics [width=88mm]{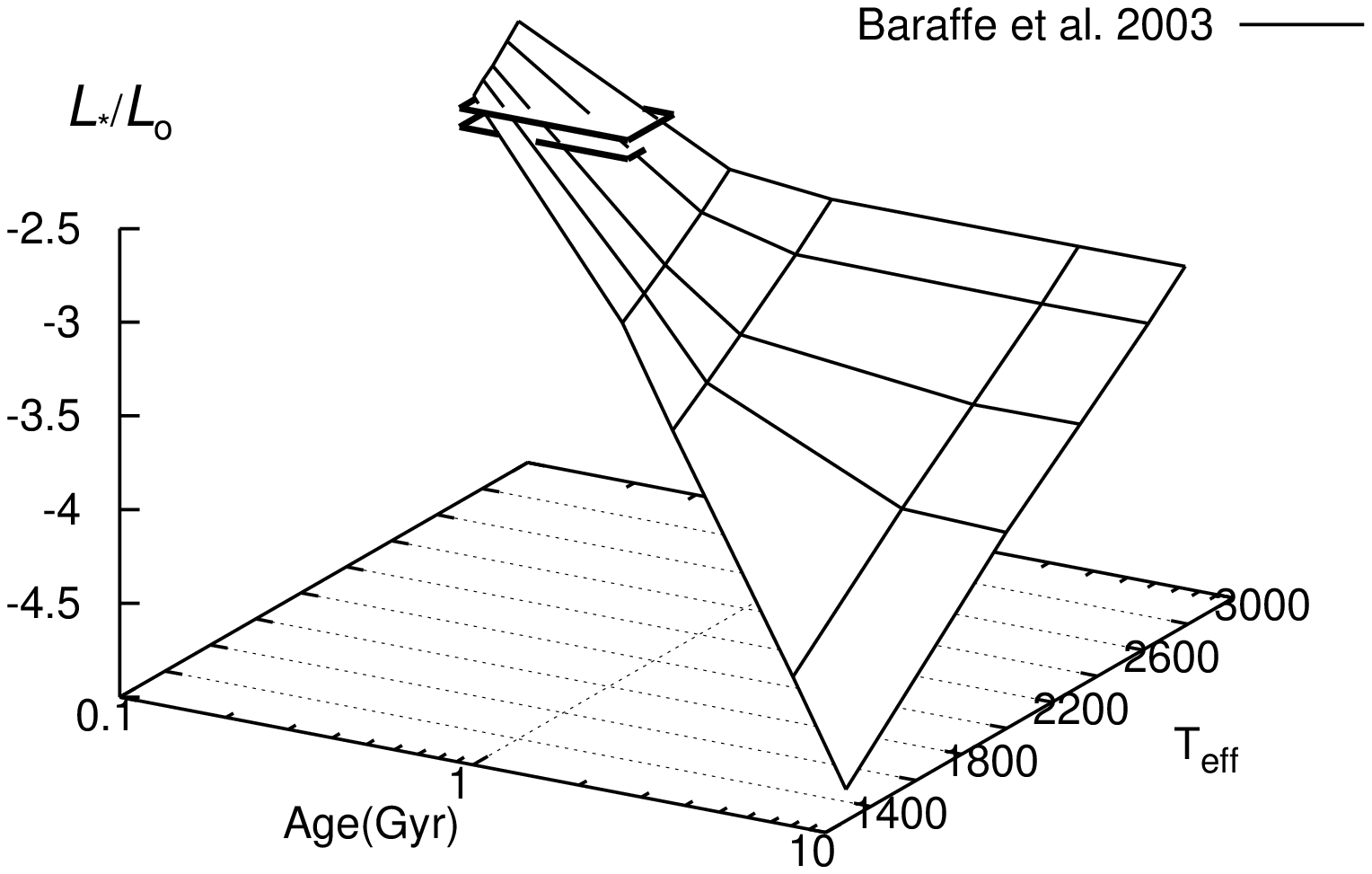}
\includegraphics [width=88mm]{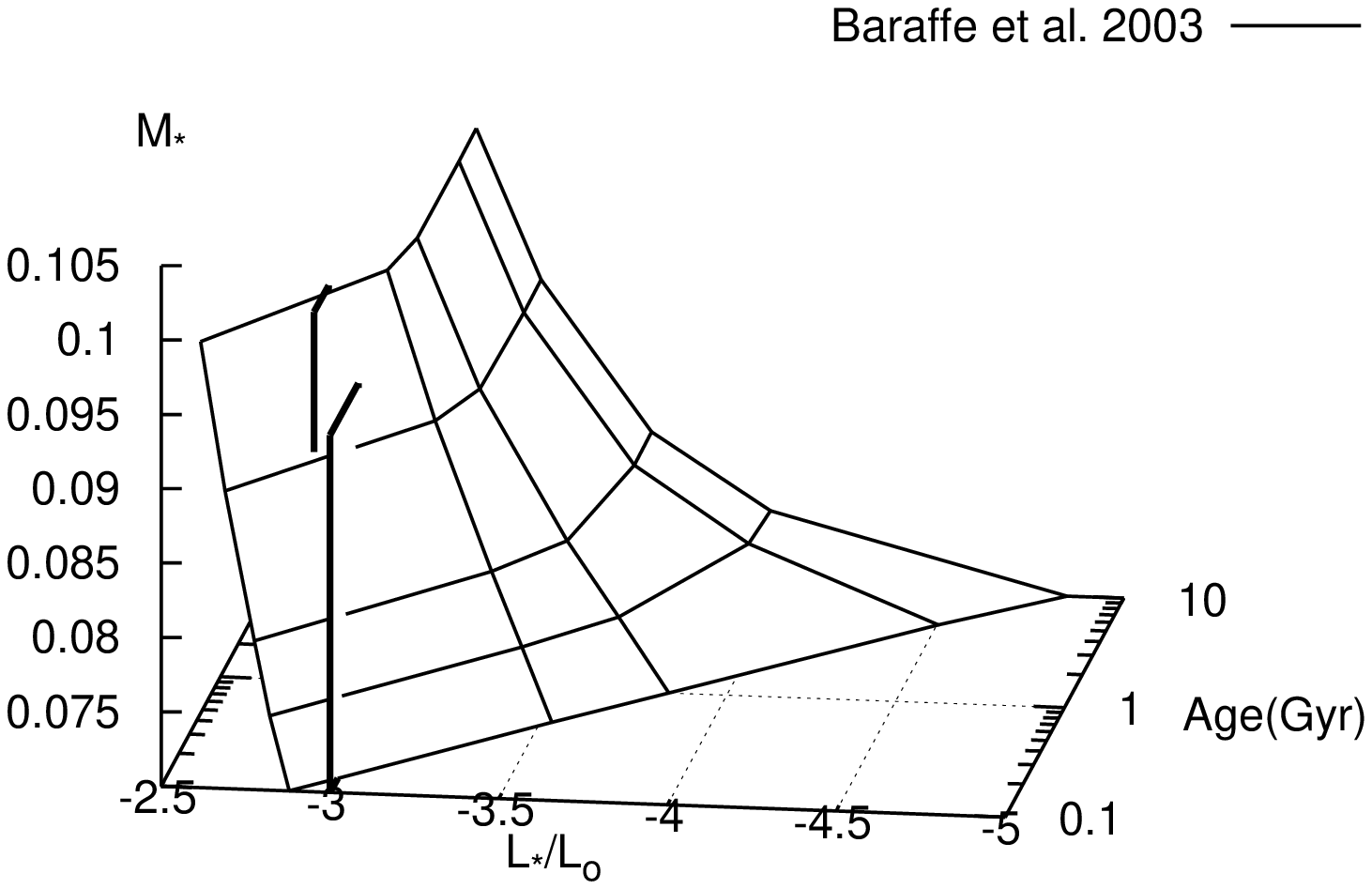}
\end{center}
\caption[]{\label{__evol} Top: location of GJ 406
in respect to
the Baraffe et al. (1998) evolutionary tracks and
\Tef = 2800, 2600 K and $L = L_*/L_\odot$ =-2.90,-3.00
planes. The planes shown on plot determine the cuboid of our errors of
\Tef and  $L$. Bottom: location of GJ 406
in respect to evolutionary tracks in coordinates $L$, $t$, $M$.
Tthe thick lines shown on plot determine the cuboid of our errors of
age, luminosity  $L_*/L_\odot$ and mass $M_*$.}
\end{figure*}

Using evolutionary models by Baraffe et al (2003),
we have estimated a mass and age for GJ 406 (Fig. \ref{__evol}).
Our measured effective
temperature \Tef =2800$\pm$100 K, and luminosity
$L$ = log $L_*/L_{\odot}$ = -2.95$\pm$0.05 agree
well with estimates of other authors (Leggett et al. 2002;
Jones et al. 2002). We know from
the lack of lithium in GJ 406 that its age must
be $>10^8$yrs (Magazzu, Mart\'{i}n \& Rebolo 1993).
Indeed, within the error bars we find that the
measured \Tef and luminosity are consistent with the Baraffe et al (2003)
 evolutionary
models for stars of \Tef = 2700 - 2900 K and $L$ = -3.0, which
complies an age of 0.1 -- 0.3 Gyr, and, respectively, $M$ = 0.07
-- 0.94 $M_{\odot}$. For $L$ = -2.9 we obtain the age 0.2 - 0.35
Gyr, and $M$ = 0.1 - 0.09 $M_{\odot}$.

\section{Discussion}

A comparison of observed and computed spectral energy distributions
provides a unique tool with which to assess the completeness and quality
of our knowledge about the structure and properties of late type dwarf
atmospheres; the physical state of their matter; opacity sources and line
lists, atmospheric temperature and pressure structure, effective temperature
scales.

Here we have modelled the spectral energy distribution
of the M6 dwarf GJ 406 from 0.4--4.9 \mum.
The optical spectrum is formed primarily by absorption of the saturated bands
of VO and TiO. In general, the response of optical fluxes to the variations of
input physical parameters is stronger than at
infrared wavelengths (see Fig.\ref{__deviat}).
However, the infrared spectrum is more sensitive to the \HHO absorption bands, with
fluxes coming from deeper atmospheric layers. Therefore, if one is to understand
both the outer and inner atmospheric structure of cool dwarfs, one must
simultaneously
account for both the optical and infrared spectra.
In general, we achieved good agreement between our theoretical
spectra and observation, although we note some problems with fits
at certain wavelengths (0.35 - 0.4 \mum, 1.3 - 1.7 \mum, these are
presumably a consequence of missing molecular opacities). Our fits
support the idea that the TiO/VO and \HHO line lists covering
these wavelength ranges are of good quality. Most probably, our
problems in the blue part of the spectrum can be solved with
proper fits to strong atomic lines located there. Then, the blue
part of the spectrum should be more affected by chromospheric like
phenomena. The detailed analysis of these and related problems is
beyond the scope of this paper. We plan to consider those in
forthcoming papers.




Our best fit age range is consistent with the age constraints from
depleted lithium. However, GJ406 has been kinematically classified as an
old disk star, and as such, one might expect its age to be greater than
$\sim$600 Myrs (the age of the Hyades, which traditionally represents an
upper age limit for the young disk population; eg. Leggett 1992). Our
model fit age range thus suggest that GJ406 is more youthful than this,
and that its old disk classification should be interpreted solely as a
kinematic description, and not as an age constraint. This is not
contradictory of course, since the dispersion in the kinematics of the
young disk population naturally places some young stars outside of the
canonical young disk UVW kinematic region.

Although the spectrum of GJ406 shows strong H$_\alpha$ emission
and activity (as summarised in Section 1), it should be noted that
one cannot use this to place any strong constraints on age. As
Gizis et al (2002) explain, one expects a significant spread in
the H$_\alpha$ emission strength of a population if the age is
less than some value that depends on the stellar colour or
spectral type (see their figure 11). For GJ406 ($V-I$ = 4.06),
this age upper limit is greater than the age of the disk, and a
high level of H$_\alpha$ emission may thus be expected for some M6
dwarfs spanning the full age of the disk. It is at least clear
that the emission properties of GJ406 are not inconsistent with
our age estimate.


\section{Acknowledgments}

The William Herschel Telescope and United Kingdom Infrared
Telescope are operated for the Particle Physics and Astronomy
Research Council (PPARC). ISO is an ESA project with the
participation of ISAS and NASA funded from member states. HST is a
NASA project funded in part by ESA. This work was partially
supported by the Royal Society and the Leverhulme Trust.
The computations were in part performed using the resources of
HiPerSPACE computing facility at UCL which is part funded by the UK
Partical Physics and Astronomy Research Council (PPARC).
We thank Maria Rosa
Zapatero Osorio for her helpful comments. We thank anonymous Referee for 
the useful
remarcs.
This research has made
use of the SIMBAD database, operated at CDS, Strasbourg, France.


\begin{thebibliography}{99}

\bibitem{}Allard F. 2005, in ``Ultra low mass star formation and evolution'',
          La Palma, 28 June -1-st July 2005, eds: A. \Magazzu and E. \Martin, in press.
\bibitem[1989]{_Anders1989_} Anders, E., Grevesse, N.,  1989, GeGoAA, 53, 197.

\bibitem{}Baraffe, I, Chabrier, G., Barman, T.S.,  Allard, F., Hauschildt, P.
2003, A\&A, 402, 711.
\bibitem{}Burrows, A., Ram, S.R., Bernath, P. 2002, ApJ., 577, 986.
\bibitem{dul2003} Dulick, M., Bauschlincher, C.W., Burrows, A.
                   2003, ApJ, 594, 651.
\bibitem{} Dobbie, P. D., Pinfield, D. J., Napiwotzki, R., Hambly, N. C.,
   Burleigh, M. R., Barstow, M. A., Jameson, R. F., Hubeny, I., 2004, MNRAS, 355, L39
\bibitem{}
        Delfosse, X., Forveille, T., Perrier, C., Mayor, M. 1998, A\&A, 331, 581.
\bibitem{} Fuhrmeister, B.,  Schmitt, J.H.M.M., Wichmann, R. 2004, A\&A, 417,
               701.
\bibitem{} Fuhrmeister, B., Schmitt, J.H.M.M., Hauschildt, P.H,
          2005, A\&A, 436, 677.
\bibitem{}Gizis, J.E.,Reid, I.N., Hawley. S.L.
       2002, 123, 3356.
\bibitem{}Goorvitch D., 1994, ApJS, 95, 535
\bibitem[1989]{_G89_} Gurvitz, L. V., Weitz, I. V., Medvedev, V.
       A. 1989, Thermodynamic properties of individual substances. Moscow.
       Science
\bibitem{} Guetter, H.H., Vrba, F.J., Henden, A.A., Luginbuhl. C.B.
    AJ, 125, 3344.
\bibitem{} Hauschildt, P. H., Allard, F., Baron, E., 1999, ApJ, 512, 377
\bibitem{} Hawarden, T. G., Leggett, S. K., Letawsky, M. B., Ballantyne, D. R., Casali, M. M.,
   2001, MNRAS, 325, 563
\bibitem{} Henry, T.J., Subasavage, J.P., Brown, M.A. et al. 2004, AJ, 128,
           2460.
\bibitem{} Hewitt, J.A., Doss, N., Zobov, N.F., Polyansky, O.L., Tennyson, J.
   2005, MNRAS, 356, 1123.
\bibitem[\protect\astroncite{Jones}{1996}]{SPU_vit97}
   Jones, H.R.A., Longmore, A.J., Jameson, R.F., Mountain, C.M., 1994,
   MNRAS, 267, 413
\bibitem{}Jones,  H. R. A., Longmore, A.. J., Allard, F., Hauschildt, P. H., 1996,
MNRAS, 280,77
\bibitem{} Jones, H.R.A., Pavlenko, Ya., Viti, S, Tennyson, J.
           2002, MNRAS, 330, 675
\bibitem{} Jones, H.R.A., Pavlenko, Ya., Viti, S, Tennyson, J.
          2th Cambridge Workshop on Cool Stars, Stellar Systems, and the
          Sun (2001 July 30 - August 3), eds. A. Brown, G.M. Harper, and T.R. Ayres,
          (University of Colorado), 2003, 899.
\bibitem{} Jones, H.R.A., Pavlenko, Ya.V., Viti, S., Barber, R.J.,
           Yakovina, L.A., Pinfield, D., Tennyson, J. 2005, 358, 105.
\bibitem[1991]{_Kirkpatrick1991_}
           Kirkpatrick, J.D., Henry, T.J., McCarthy, D.W., 1991, ApJS, 77, 417
\bibitem[1999]{_VALD2_} Kupka, F., Piskunov, N., Ryabchikova, T.
           A., Stempels, H. C., Weiss, W. W. 1999, A\&A. Suppl. Ser. 138,
           119.
\bibitem{}Kurucz, R.L. 1970, SAO Spec. Rept. N 309, 291
\bibitem{}Kurucz, R,L. 1993, CDROMs 1-22, Harvard-Smisthonian
                                                       Observatory.
\bibitem{}Kurucz, R,L. 1999, http://kurucz.harvard.edu
\bibitem{} Langhoff, S.R., 1997, ApJ, 481, 1007.
\bibitem{}Leggett, S. K., 1992, ApJS, 82, 351
\bibitem{}Leggett, S. K., Allard, F., Hauschildt, P H., 1998, ApJ, 509, 836
\bibitem{}Leggett, S.K., Allard, F., Dahn, C., Hauschildt, P.H.,
          Kerr, T.H., Rayner, J., 2000, ApJ, 535, 965
\bibitem{}Lettett, S.K., Golimowski, D.A.,Fan, X. et al. 2002, ApJ, 564, 452
\bibitem{}Lyubchik, Yu., Pavlenko, Ya. 2001, KFNT, 17, 017.
\bibitem{}\Magazzu, A., Mart\'{i}n, E. L., Rebolo, R., 1993, ApJ, 404, L17
\bibitem{}Marino, A., Micela, G., Peres, G., 2000, A\&A, 353, 177
\bibitem{}Mart\'{i}n, E. L., Basri, G., Delfosse, X. \& Forveille, T, 1997, A\&A, 327, L29.
\bibitem[2004]{_m3_} Mohanty, S., Basri, G. 2003, ApJ, 583, 451.
\bibitem[2004]{_m4_} Mohanty, S., Basri, G., Javardhana, R. et al.
               2004, ApJ, 609, 854.
\bibitem[1998]{_PS1998_} Partrige, H., Schwenke, D.J. 1997,
                          Chem. Phys. 106, 4618.
\bibitem[1889]{Plez1989_} Plez, B., 1998, A\&A, 337, 495.
\bibitem[1997]{pav97} Pavlenko, Ya. 1997, Astrophys. Space Sci., 253, 43
\bibitem[1998]{pav98} Pavlenko, Ya.\,V. 1998, Astron. Rept., 42, 501
\bibitem[2000]{pav00} Pavlenko, Ya. 2000, Astron. Rept., 44, 219
 \bibitem[2001]{pav01} Pavlenko, Ya. 2001, Astron. Rept., 45, 144
 \bibitem[2001]{pav02} Pavlenko, Ya. 2002, Astron. Rept., 46, 567
\bibitem[Pavlenko \& Jones(2002)]{pav-jones} Pavlenko, Ya. V., Jones, H. R. A.,
         2002, A\&A, 396, 967
\bibitem[]{} Pirzkal, N., \& Freudling, W. 1998,
NICMOS and the VLT: A New Era
of High Resolution Near Infrared Imaging and Spectroscopy, Pula,
Sardinia, Italy, June 26-27th 1998 ESO Conference and Workshop Proceedings 55,
Wolfram Freudling and Richard Hook eds.,p. 55.
\bibitem[]{}
       Reid, I.N., Kirkpatrick, J.D., Liebert, J., et al. 1999, ApJ, 521, 613.
\bibitem[]{}
         Schmitt, J. H. M. M., Wichmann, R. 2001, Nature, 412, 508.
\bibitem[]{}
          Schmitt, J. H. M. M.,    and Liefke, C. 2004, A\&A, 417,651.
\bibitem[]{} Schwenke, D., 1998, Faraday Discuss., 109, 321.
\bibitem[1973]{tsuji73} Tsuji, T. 1973, A\&A, 23, 411
\bibitem[1997]{_Tsuji1997_}
                    Tsuji, T.; Ohnaka, K.; Aoki, W. 1997. ASP
                     Conference Series, 124,  ed. H. Okuda; T.
                     Matsumoto, T. Rollig, 91
\bibitem{} van Altena, W.F., Lee, J.T., Hoffleit,E.D. 1995, The general
    catalogue of trigomometric stellar parameters (4-th rev.ed: Schenrctady:
    L.Davis).
\bibitem{} Vidler, M., Tennyson, J. 2000, Journ. of Chem. Pgysics,
                               113, 21.
\bibitem[\protect\astroncite{Unsold}{1999}]{Unsold55}
Uns\"old, A. Physics der Sternatmospharen, 1955. Springer.


\end{thebibliography}
\end{document}